\documentclass[fleqn,usenatbib]{mnras}

\usepackage{newtxtext,newtxmath}

\usepackage[T1]{fontenc}
\usepackage{ae,aecompl}
\usepackage{csquotes}

\usepackage{graphicx}	
\usepackage{amsmath}	
\usepackage{float}
\usepackage{multicol}
\usepackage{bm}
\usepackage{subfigure}
\usepackage{relsize}
\usepackage{anyfontsize}
\usepackage[dvipsnames]{xcolor}
\usepackage{url}
\usepackage{enumerate}

\title[Thermal luminosity degeneracy of neutron stars]{Thermal luminosity degeneracy of magnetized neutron stars with and without hyperon cores}

\author[F. Anzuini, A. Melatos, C. Dehman, D. Vigan\`o,  J. A. Pons ]{
F. Anzuini,$^{1,2,3}$\thanks{E-mail: filippo.anzuini@monash.edu}
A. Melatos$^{1, 4}$\thanks{E-mail: amelatos@unimelb.edu.au}, C. Dehman$^{5, 6}$, D. Vigan\`o$^{5, 6, 7}$, J. A. Pons$^{8}$
\\
$^{1}$School of Physics, University of Melbourne, Parkville, Victoria 3010, Australia\\
$^{2}$School of Physics and Astronomy, Monash University, Victoria 3800, Australia\\
$^{3}$OzGrav: The ARC Centre of Excellence for Gravitational Wave Discovery, Clayton, Victoria 3800, Australia\\
$^{4}$Australian Research Council Centre of Excellence for Gravitational Wave Discovery (OzGrav), University of Melbourne, \\
 \  Parkville, Victoria 3010, Australia\\
$^{5}$Institute of Space Sciences (IEEC-CSIC), Campus UAB, Carrer de Can Magrans s/n, 08193, Barcelona, Spain\\
$^{6}$Institut d'Estudis Espacials de Catalunya (IEEC), Carrer Gran Capità 2–4, 08034 Barcelona, Spain\\
$^{7}$Institute of Applied Computing \& Community Code (IAC3), University of the Balearic Islands, Palma, 07122, Spain\\
$^{8}$ Departament de Física Aplicada, Universitat d'Alacant, 03690 Alicante, Spain \\
}

\date{Accepted XXX. Received YYY; in original form ZZZ}

\pubyear{2021}

\begin{document}
\label{firstpage}
\pagerange{\pageref{firstpage}--\pageref{lastpage}}
\maketitle

\begin{abstract}
The dissipation of intense crustal electric currents produces high Joule heating rates in cooling neutron stars. Here it is shown that Joule heating can counterbalance fast cooling, making it difficult to infer the presence of hyperons (which accelerate cooling) from measurements of the observed thermal luminosity $L_\gamma$. Models with and without hyperon cores match $L_{\gamma}$ of young magnetars (with poloidal-dipolar field $B_{\textrm{dip}} \gtrsim 10^{14}$ G at the polar surface and $L_{\gamma} \gtrsim 10^{34}$ erg s$^{-1}$ at $t \lesssim 10^5$ yr) as well as mature, moderately magnetized stars (with $B_{\textrm{dip}} \lesssim 10^{14}$ G and  $10^{31} \ \textrm{erg s}^{-1} \lesssim L_{\gamma} \lesssim 10^{32}$ erg s$^{-1}$ at $t \gtrsim 10^5$ yr). In magnetars, the crustal temperature is almost independent of hyperon direct Urca cooling in the core, regardless of whether the latter is suppressed or not by hyperon superfluidity. The thermal luminosities of light magnetars without hyperons and heavy magnetars with hyperons have $L_{\gamma}$ in the same range and are almost indistinguishable. Likewise, $L_{\gamma}$ data of neutron stars with $B_{\textrm{dip}} \lesssim 10^{14}$ G but with strong internal fields are not suitable to extract information about the equation of state as long as hyperons are superfluid, with maximum amplitude of the energy gaps of the order $\approx 1$ MeV.

\end{abstract}




\section{Introduction}
\label{sec:Introduction}
Timing measurements of neutron stars show that their inferred magnetic field spans a wide range of strengths, from $\sim 10^8$ G in millisecond pulsars \citep{Backer_1982, Boriakoff_1983, Lyne_1987, Manchester_2017, Arzoumanian_2018} up to $\sim 10^{15}$ G in magnetars \citep{Mazets_1979, Mazets_1981, Gavriil_2002, Vigano_2013, Vogel_2014, Olausen_2014, Mereghetti_2015, Kaspi_2017}.
 Although the magnetic field configuration of neutron stars at birth is unknown \citep{Duncan_1992, Thompson_1993, Spruit_2008}, numerous authors have studied the possible initial magnetic field configuration consistent with MHD-equilibrium \citep{Braithwaite_2006, Ciolfi_2009, Lander_2009, Ciolfi_2013} and the long-term evolution of both crust-confined or core-threading topologies \citep{Gourgouliatos_2013, Vigano_2013, Wood_2015, Gourgouliatos_2016, Elfritz_2016, Igoshev_2021, DeGrandis_2021}. These initial configurations are likely over-simplified. For example, crustal confinement is not guaranteed, and it is unclear how the twisted torus magnetic configuration (one of the most common initial topologies employed in numerical studies) would be produced. As a matter of fact, recent magnetohydrodynamic simulations of the magnetorotational instability in core-collapse supernovae  \citep{Aloy_2021, Reboul-Salze_2021} suggest a different and more complex picture, in which the magnetic energy of the protoneutron star spreads over a wide range of spatial scales. Such simulations find that most of the magnetic energy is contained in small or medium-scale size magnetic structures, both for the dominant toroidal components, and the weaker poloidal components. 
 
 The magnetic evolution of a neutron star is interlinked with its thermal evolution and hence with its composition, which depends on the equation of state \citep{Aguilera_2008, Vigano_2013, Pons_2019, Dehman_2020,  Vigano_2021, Igoshev_2021, DeGrandis_2021, Anzuini_2021}. In particular, the presence of exotic species such as hyperons accelerates cooling via direct Urca processes \citep{Prakash_1992, Yakovlev_2001}. In turn, the magnetic field causes anisotropic heat transport across and along the magnetic field lines, and the internal layers are heated up by the dissipation of the electric currents that sustain the field (Joule heating).

In this paper, we show that Joule heating hides the effect of fast cooling on the observed thermal luminosity $L_\gamma$. We extend the results presented in \cite{Aguilera_2008} by simulating the self-consistent evolution of the magnetic field, including Ohmic dissipation and the generation of small scales by the action of the Hall drift. Moreover, we consider hyperons in the core, unlike in \cite{Aguilera_2008}, and focus on the cooling effect of hyperon direct Urca. Accelerated direct Urca cooling is an important signature of the presence of hyperons, so Joule heating complicates the link between cooling curves, internal composition, and ultimately the equation of state (EoS). We find that the magneto-thermal evolution of light stars without hyperon cores resembles the evolution of heavy stars with hyperon cores, if the crustal field is sufficiently strong. The thermal power produced by Joule heating due to magnetic field decay dominates the thermal evolution of the crust, with less influence from neutrino cooling in the core, so that the temperatures of the crust and core \enquote{decouple}, i.e. they evolve approximately independently \citep{Kaminker_2006, Kaminker_2007}. Part of the additional heat is dispersed via neutrino emission, and part of it is transported via thermal conduction to the surface, increasing $L_\gamma$. We show that high Joule heating rates affect the interpretation of $L_{\gamma}$ data in terms of light models without hyperons and heavy models with hyperons for young magnetars ($t \lesssim 10^5$ yr), with surface value at the pole of the poloidal-dipolar field $B_{\textrm{dip}} \gtrsim 10^{14}$ G, and mature stars ($t \gtrsim 10^5$ yr) with $B_{\textrm{dip}} \lesssim 10^{14}$ G. 

We study the magneto-thermal evolution of models with or without concentrations of hyperons in their cores with the updated version of the two-dimensional, axisymmetric magneto-thermal code developed by the Alicante group \citep{Aguilera_2008, Pons_2009, Vigano_2012, Vigano_2013, Pons_2019, Dehman_2020, Vigano_2021}, recently adapted to study hyperon stars \citep{Anzuini_2021}. We employ the GM1A EoS \citep{Gusakov_2014}, based on the $\sigma\omega\rho\phi\sigma^*$ model of nucleon, lepton and hyperon matter. The model is fitted to hypernuclear data, and predicts that only $\Lambda$ and $\Xi^-$ hyperons appear in dense matter. Other hyperon species, such as $\Sigma^-$ hyperons, do not appear in the allowed density range. We extend the results found in \cite{Anzuini_2021} by considering both crust-confined and core-extended initial magnetic field configurations. Neutron star models obtained with the GM1A EoS cool down rapidly due to the activation of both nucleonic and hyperonic direct Urca emission. If neutrons are paired in a large fraction of the stellar core, internal heating is required to match $L_{\gamma}$ data \citep{Anzuini_2021}. Among the possible heating mechanisms \citep{Alpar_1984, Shibazaki_1989, Fernandez_2005, Pons_2007a, Vigano_2013, Hamaguchi_2019, Pons_2019}, Joule heating can supply the necessary thermal power to reconcile theoretical cooling rates and $L_{\gamma}$ observations.

This paper is organized as follows. Section \ref{sec:Model} describes the theoretical framework adopted to simulate the magneto-thermal evolution of neutron stars. It introduces the heat diffusion and magnetic induction equations, as well as the microphysics input. In Section \ref{sec:Observations} we calculate $L_\gamma$ versus time for a selection of representative magneto-thermal models. The corresponding surface temperatures are studied in Section \ref{sec:Surface_temp}.

\section{Stellar model}
\label{sec:Model}

In this section we outline the ingredients of the model describing the star's magneto-thermal evolution. We introduce the heat diffusion and the magnetic induction equations in Section \ref{sec:equations}, and the initial conditions for the magneto-thermal evolution in Section \ref{sec:initial}. The microphysics input (e.g. superfluid model and neutrino emissivity) is described in Section \ref{sec:miscrophysics}. Section \ref{sec:mass_models} contextualizes the mass range studied in this paper in terms of available neutron star data. Degeneracies arise when comparing the model output (e.g. cooling curves) with observations of $L_\gamma$, as described in Section \ref{sec:Internal_heating}.

The magneto-thermal evolution of neutron stars is studied assuming that the space-time metric is given by the Schwarzschild metric. Deviations from spherical symmetry related to the temperature and magnetic field are neglected \citep{Pons_2019}.

\subsection{Heat diffusion, magnetic induction}
\label{sec:equations}

The internal temperature evolves via the heat diffusion equation \citep{Aguilera_2008, Pons_2009}

\begin{eqnarray}
    c_\textrm{V}e^{\Phi}\frac{\partial T}{\partial t} + \boldsymbol{\nabla}\cdot(e^{2\Phi} \boldsymbol{F}) = e^{2\Phi}(Q_{\textrm{J}} - Q_{\nu}) \ .
    \label{eq:thermal_evolution}
\end{eqnarray}
In Eq. \eqref{eq:thermal_evolution}, the heat capacity per unit volume of nucleons, leptons and hyperons is denoted by $c_\textrm{V}$. The internal, local temperature and the dimensionless gravitational potential are $T$ and $\Phi$ respectively, and the differential operator $\nabla$ includes the metric factors. The heat flux $\boldsymbol{F}$ reads $\boldsymbol{F} = - e^{-\Phi} \hat{k} \cdot \boldsymbol{\nabla}(e^{\Phi}T)$, where $\hat{k}$ denotes the thermal conductivity tensor \citep{Potekhin_2001, Potekhin_2003}, while $Q_{\textrm{J}}$ and $Q_{\nu}$ denote respectively the Joule heating rate per unit volume and neutrino emissivity per unit volume. 

Given the high thermal conductivity of the core, the latter becomes isothermal a few decades after the neutron star birth. The crust relaxes slower thermally, during a period that typically lasts $t \sim 10^2$ yr, depending on the thermal conductivity, heat capacity of the crust layers and whether neutrons are superfluid \citep{Lattimer_1994, Gnedin_2001}. During the thermal relaxation stage the crust temperature is higher than the core temperature, and the thermal luminosity of the star does not reflect the thermal evolution of the core. The relaxation stage ends when the \enquote{cooling wave} \citep{Gnedin_2001} propagating from the core reaches the stellar surface, and the thermal luminosity drops by orders of magnitude (depending, among other factors, on the presence of superfluid phases).

We solve the heat-diffusion equation everywhere in the stellar interior, except in the outer envelope. The typical time-scales in the envelope are shorter than in the deeper layers, requiring a smaller time-step and increasing the computational cost. Instead, we rely on an effective relation between the internal temperature at the bottom of the outer envelope $T_\textrm{b}$ and the surface temperature $T_\textrm{s}$. The latter is obtained from the $T_{\textrm{s}}$ -- $T_{\textrm{b}}$ relation employed in \cite{Potekhin_2015}, \cite{Vigano_2021} and \cite{Anzuini_2021}. The $T_{\textrm{s}}$ -- $T_{\textrm{b}}$ relation depends on the magnetic field \citep{Potekhin_2015}; see also the discussion in \cite{Anzuini_2021}. In the following, we assume that the outer envelope is composed of iron.

The magnetic field $\mathbf{B}$ evolves according to the magnetic induction equation, which in the crust reads 
\begin{eqnarray}
\frac{\partial \boldsymbol{B}}{\partial t} = -\boldsymbol{\nabla}\times \Big[ \frac{c^2}{4 \pi \sigma_e} \boldsymbol{\nabla}\times (e^{\Phi}\boldsymbol{B}) + \frac{c}{4 \pi e n_e}[\boldsymbol{\nabla} \times (e^{\Phi}\boldsymbol{B})] \times \boldsymbol{B} \Big] \ ,
\label{eqn:induction}
\end{eqnarray}
where $c$ is the speed of light, $\sigma_e$ is the temperature- and density-dependent electrical conductivity, $e$ is the elementary electric charge and $n_e$ is the electron number density.
The first term is the Ohmic (dissipative) term and the second is the nonlinear Hall term. 

As the temperature drops due to neutrino emission, the thermal and electric conductivities increase and become temperature-independent for sufficiently low temperatures \citep{Aguilera_2008}, gradually decreasing the Ohmic dissipation rate. At the same time, the decay of the magnetic field is enhanced by the Hall term in the magnetic induction equation, because although the Hall term does not directly dissipate magnetic energy, it produces small-scale magnetic structures, where Ohmic dissipation is enhanced. Furthermore, the Hall drift tends to push the electric currents toward the crust-core boundary, where they may be dissipated more efficiently by the presence of impurities and pasta phases \citep{Pons_2013, Vigano_2013}, producing higher Joule heating rates. As the magnetic field evolves, the thermal conductivity along and across the magnetic field lines changes, affecting the local temperature. Hence, the magnetic evolution influences the thermal evolution and vice versa.

The evolution of the magnetic field in the core is more uncertain due to its multifluid nature and the occurrence of proton superconductivity. There may be regions with protons in the normal phase or in the superconducting phase, the latter being of type-II \citep{Baym_1969, Sedrakian_2019} or type-I (leading to magnetic field expulsion due to the Meissner effect). Recent calculations \citep{wood_2020} predict phase coexistence in mesoscopic regions (larger than the flux tubes, but smaller than the macroscopic length-scales), introducing several length and time-scales into the problem. When superconductivity is neglected, the typical time-scales for Ohmic dissipation and Hall advection exceed the cooling time-scales, so that the magnetic field undergoes little change in the stellar core \citep{Elfritz_2016, Dehman_2020, Vigano_2021}. The inclusion of ambipolar diffusion could partially speed up the dynamics under certain conditions \citep{castillo_2020}. A more consistent approach including hydrodynamic effects in the superfluid/superconducting core could substantially accelerate the evolution, as recently discussed in terms of estimated time-scales by \cite{Gusakov_2020} (see also references therein). Moreover, as noted above, an initial complex magnetic topology can also reduce the typical length- and time-scales, compared to the usually assumed purely dipolar fields. 

In the simulations reported in this work, the induction equation in the crust includes both the Ohmic and Hall terms. In the core, only the Ohmic term is included, so that the core magnetic field is frozen, since the typical diffusion timescale in the core is larger than the typical age of isolated neutron stars studied here.

\subsection{Initial conditions}
\label{sec:initial}

\begin{table}
\centering
\caption{\label{tab:Topology_Table_crust} Crust-confined initial magnetic configurations for a star with $M = 1.8 \ M_{\odot}$. $B_{\textrm{dip}}$ is the surface field strength at the pole of the dipolar-poloidal component. $E^{\textrm{T}}_{\textrm{mag}}$ is the magnetic energy stored in the toroidal component, and $E_{\textrm{mag}}$ denotes the total magnetic energy. The number of poloidal multipoles in the crust is denoted by $l_{\textrm{pol}}$.}
\begin{tabular}{cccccc}
\hline
\hline
Config.&
\multicolumn{1}{c}{$B_{\textrm{dip}}$}&
\multicolumn{1}{c}{$E^{\textrm{T}}_{\textrm{mag}}/ E_{\textrm{mag}}$}&
\multicolumn{1}{c}{$l_{\textrm{pol}}$}\\
\hline

A1 &$1.0 \times 10^{13}$ G & $93 \%$ & $1$ \\
A2 & $5.0 \times 10^{13}$ G & $ 35 \%$ & $1$ \\
A3 & $1.0 \times 10^{14}$ G & $ 35 \%$ & $1$\\
A4 & $1.0 \times 10^{15}$ G & $ 0.5 \%$ & $1$ \\
A5 & $2.0 \times 10^{15}$ G & $0$ & $1$ \\
A1m$_{2,2}$ & $1.0 \times 10^{13}$ G & $77 \%$ & $2$  \\
A1m$_{3,3}$ & $1.0 \times 10^{13}$ G & $51 \%$ & $3$  \\
A1m$_{4,4}$ & $1.0 \times 10^{13}$ G & $29 \%$ & $4$ \\
\hline
\end{tabular}
\end{table}

\begin{table}
\centering
\caption{\label{tab:Topology_Table_core} Core-extended initial magnetic configurations. The quantities $B_{\textrm{dip}}$, $E^{\textrm{T}}_{\textrm{mag}}$, $E_{\textrm{mag}}$ and $l_{\textrm{pol}}$ are defined as in Table \ref{tab:Topology_Table_crust}. In the B1, B2 and C1 configurations the initial toroidal field is confined to an equatorial torus in the core.}
\begin{tabular}{cccccc}
\hline
\hline
Config.&
\multicolumn{1}{c}{$B_{\textrm{dip}}$}&
\multicolumn{1}{c}{$E^{\textrm{T}}_{\textrm{mag}}/ E_{\textrm{mag}}$}&
\multicolumn{1}{c}{$l_{\textrm{pol}}$}\\
\hline

B1 &$1.0 \times 10^{13}$ G & $50 \%$ & $1$ \\
B2 & $1.0 \times 10^{14}$ G & $50 \%$ & $1$\\
C1 & $1.0 \times 10^{14}$ G & $42 \%$ & $1$\\
C2 & $2.0 \times 10^{14}$ G & $ 0$ & $1$\\
C1m$_{2,0}$ & $1.0 \times 10^{14}$ G & $ 0 $ & $2$\\
\hline
\end{tabular}
\end{table}

The evolution is independent of the initial internal temperature (if the latter is sufficiently high), and we typically adopt a temperature of $10^{10}$ K \citep{Yakovlev_1999, Page_2004, Page_2006, Potekhin_2015}.

The magnetic fields of neutron stars may be sustained by electric currents both in the crust and in the core. In particular, in the crust the currents can produce small-scale magnetic fields that enhance Joule heating via the Hall cascade \citep{Gourgouliatos_2019,Brandenburg_2020}.
In this work we consider various possible initial magnetic field configurations (listed in Tables \ref{tab:Topology_Table_crust} and \ref{tab:Topology_Table_core}). The two main categories are the following. (i) \textit{Crust-confined fields.} The radial magnetic field component vanishes at the crust-core interface, while the latitudinal ($B_{\theta}$) and toroidal ($B_{\phi}$) components are different from zero. (ii) \textit{Core-threading fields.} At the crust-core interface the radial component of the magnetic field is $B_r \neq 0$, and the magnetic field lines penetrate into the core. In both cases, at the surface the magnetic field is matched continuously with the potential solution of a force-free field (i.e. the electric currents do not leak into the magnetosphere).

In Table \ref{tab:Topology_Table_crust} we list the crust-confined magnetic field configurations considered in this work. From a computational point of view, there is limited capability to follow numerically the rich dynamics of small-scale magnetic fields in the crust; however, the configurations studied here may reproduce typical Joule heating rates of more realistic, small-scale crustal fields. We vary the ratio of the magnetic energy stored in the toroidal component ($E^{\textrm{T}}_{\textrm{mag}}$) and the total magnetic energy ($E_{\textrm{mag}}$), as well as the number of poloidal multipoles $l_{\textrm{pol}}$.
Crust-confined magnetic fields generate typically higher Joule heating rates than core-extended fields, for similar values of the total magnetic energy. As a matter of fact, in the crust-confined case all currents are forced to circulate in the crust, where the resistivity is orders of magnitude larger than in the core.

We consider two families of core-extended configurations (listed in Table \ref{tab:Topology_Table_core}). The first family includes the B1 and B2 configurations (studied for example by \cite{Dehman_2020} and \cite{Vigano_2021}), where the electric currents that sustain the magnetic field reside exclusively in the core. The corresponding Joule heating is typically lower than crust-confined configurations for two reasons. First, the resistivity in the core is low, so that Joule heating is lower than in the crust. Second, any additional heat produced via Joule heating in the core is carried away by neutrinos. The second family includes configurations with both crustal and core electric currents. To mimic the total currents in both the crust and core in neutron stars, in the C1, C2 and C1m$_{2,0}$ configurations we assume the existence of large-scale poloidal-dipolar fields threading the core, plus crustal fields. For example, in the C1 configuration, a large-scale dipolar-poloidal field threads the star, sustained by currents in the core. Additionally, there is a crustal dipolar-poloidal field, sustained by crustal currents. The azimuthal field is almost entirely confined to a torus in the core. In the C1m$_{2,0}$ configuration, there is a core-threading large-scale poloidal dipole sustained by electric currents in the core, plus an additional poloidal field with two multipoles in the crust. The core-extended configurations with crustal and core electric currents reproduce qualitatively similar Joule heating rates to the ones expected from small-scale, crustal fields, attained by the crust-confined configurations in Table \ref{tab:Topology_Table_crust}.

\subsection{Microphysics input}
\label{sec:miscrophysics}
Neutron star models calculated with the GM1A EoS include concentrations of nucleons, leptons and hyperons ($npe\mu Y$ matter, where $Y$ denotes hyperonic species). In particular, the EoS is fitted to modern hypernuclear data (e.g. \cite{Millener_1988, Schaffner_1994, Takahashi_2001, Weissenborn_2012}, see \cite{Gusakov_2014}), and predicts that only the $\Lambda$ and $\Xi^{-}$ hyperons appear in dense matter, while $\Sigma^-$ hyperons are absent because their potential in dense nuclear matter is repulsive. Below we list concisely the microphysics input in our simulations, such as heat capacity, neutrino emissivity, and thermal and electric conductivities, emphasizing the novelties compared to the last version of the code \citep{Vigano_2021, Anzuini_2021}.

\begin{itemize}
    \item \textit{Heat capacity.} We include the contribution to the heat capacity of $npe\mu Y$ matter \citep{Yakovlev_1999} as well as the contribution of ions in the crustal rigid lattice. 
    \item \textit{Neutrino emission.} In the core, neutrinos are produced via nucleonic and hyperon direct Urca reactions, Cooper pair breaking and formation processes, neutrino bremsstrahlung and modified Urca. We implement the in-medium corrections to the modified Urca process emission rates in \cite{Shternin_2018}, where the enhancement factors are calculated only for the neutron branch (Eqs. (8) and (9) in \cite{Shternin_2018}). We apply the same formulae to the proton branch as well, which we interpret as upper limits to the in-medium corrections\footnote{We check that the correction factors implemented in our code reproduce the results in \cite{Shternin_2018} for the BCPM EoS \citep{Sharma_2015}. }. Such corrections make a negligible impact on the cooling curves in our case, given the superfluid model adopted (see below) and the activation of direct Urca cooling processes. For the crust, we match the GM1A EoS with the SLy4 EoS \citep{Douchin_2001}, including the crustal neutrino emission processes considered in \cite{Anzuini_2021}. We note that neutron star models obtained with the GM1A EoS cool down fast via nucleonic direct Urca (which is active for $M \geq 1.1 \ M_{\odot}$). For $M \gtrsim 1.49 \ M_{\odot}$ the hyperon direct Urca involving protons and $\Lambda$ hyperons is triggered, and for $M \gtrsim 1.67 \ M_{\odot}$ the hyperon direct Urca involving $\Xi^{-}$ and $\Lambda$ hyperons activates. 
    \item \textit{Electric and thermal conductivities.} The conductivities depend on density and temperature and vary by orders of magnitude in the crust and core regions. In our simulations, the thermal conductivity is a tensor because of anisotropic heat transport caused by the magnetic field \citep{Potekhin_2001, Potekhin_2003}, with components parallel and perpendicular to the magnetic field lines. We do not include contributions of nucleons, muons or hyperons to the electrical conductivities due to their lower mobility with respect to electrons.
    \item \textit{Superfluid phases.} Nucleons and hyperons can be superfluid, suppressing both the heat capacity and most channels of neutrino production (only partially compensated by the Cooper pair breaking and formation neutrino channel). As in \cite{Anzuini_2021}, we assume that neutrons pair in the singlet channel in the crust (\enquote{SFB} model in \cite{Ho_2015}) and in the triplet channel in the core (\enquote{c} model in \cite{Page_2004}, see also \cite{Yanagi_2020}). Protons pair in the singlet channel throughout the stellar core (\enquote{CCDK} model \citep{Ho_2015}). We use the parameters reported in Appendix A in \cite{Anzuini_2021} for singlet pairing of hyperon species, reproducing similar gaps to the ones calculated by \cite{Raduta_2018}. We also neglect the occurrence of nucleon-hyperon superfluid phases arising from the interaction of nucleons and hyperons \citep{Zhou_2005, Nemura_2009, Haidenbauer_2020, Sasaki_2020, Kamiya_2022}. The study of the magneto-thermal evolution with different hyperon superfluid gaps, obtained for example from lattice quantum chromodynamics simulations \citep{Aoki_2008, Aoki_2012, Hatsuda_2018, Sasaki_2020, Kamiya_2022} is left for future work.
\end{itemize}

\subsection{Mass models}
\label{sec:mass_models}
We study the thermal luminosity of hyperon and non-hyperon stars by comparing the magneto-thermal evolution of light-mass models ($M = 1.3 \ M_{\odot}$) and massive models with hyperon concentrations in the core ($M = 1.8 \ M_{\odot}$).  Given the large nucleon and hyperon gaps, the thermal luminosity of low-mass stars with $M = 1.3 \ M_{\odot}$ is similar to models with masses in the range $1.1 \ M_{\odot} \lesssim M \lesssim 1.4 \ M_{\odot}$ (see \cite{Anzuini_2021}). The same applies to the thermal luminosities of high-mass stars with $M = 1.8 \ M_{\odot}$, which are similar to models with masses in the range with $1.5 \ M_{\odot} \lesssim M \lesssim 1.8 \ M_{\odot}$ \citep{Anzuini_2021}.

We emphasize that neutron stars are commonly found with masses of the order of $M \approx 1.3 \ M_{\odot}$, while heavy stars are less common. Massive stars may form in merger events (if the remnant does not collapse into a black hole) \citep{Fryer_2015, Mandel_2020, Ruiz_2021}, or due to matter accreted over long time-scales (up to $\approx 0.1 \ M_{\odot}$ in roughly $10$ Gyr, in the optimal scenario) \citep{Chevalier_1989, Kiziltan_2013} for example. Some mass measurements obtained via X-ray and optical observations of neutron stars in binary systems with white dwarfs fall in the range $M \gtrsim 1.6 \ M_{\odot}$ \citep{Kiziltan_2013, Alsing_2018}. Heavy stars formed via merging or accretion may have inhomogeneous internal temperatures and complex magnetic field configurations, far from the initial conditions commonly employed in the literature of neutron star cooling. For the purpose of this work (i.e. the study of the $L_\gamma$ degeneracy between low-mass and high-mass models), we adopt the standard initial conditions employed by several authors \citep{Yakovlev_1999, Page_2004, Potekhin_2018, Raduta_2018, Raduta_2019}, bearing in mind that the magneto-thermal evolution of heavy stars likely requires more realistic temperature and magnetic field configurations initially.

\subsection{Internal heating}
\label{sec:Internal_heating}

In principle, it should be possible to constrain the internal composition of a neutron star and hence the EoS of dense matter by comparing the output of cooling simulations, specifically $L_\gamma$ as a function of the stellar age, with optical and X-ray measurements of $L_\gamma$ \citep{Vigano_2013, Potekhin_2020}. In practice, there are several scenarios where the task is complicated by internal heating. 

Consider an internal heating mechanism in the crust. When the thermal power supplied by the source is high enough, the local temperature increases and becomes almost independent of the influence of the neutrino emission processes deep in the core, so that the crust and core are thermally decoupled \citep{Kaminker_2006, Kaminker_2007, Kaminker_2009, Kaminker_2014}. We emphasize that decoupling in this context means that the crust and core temperatures evolve approximately independently. It does not mean that the crust and core are thermally insulated; there is still a heat flux between the crust and core. One key role is played by the location where the additional thermal power is supplied \citep{Kaminker_2006, Anzuini_2021}. If the additional heat is supplied at the bottom of the crust, it is transported via thermal conduction to the core, where it is easily lost via neutrino emission processes. As a result, the crust and core are thermally coupled, and the star cools down faster. On the other hand, if the heater deposits heat close to the outer envelope, it increases the local temperature, a fraction of the heat is transported via thermal conduction to the surface (increasing the surface temperature and hence $L_\gamma$), and a fraction makes its way into the core, where it is lost by neutrino emission processes. If the heating rate is sufficiently high, the local temperature in the crust is dominated by the heater, and the thermal evolution of the crust decouples from the core. In this scenario, it is challenging to ascertain whether the observed $L_{\gamma}$ is the result of fast cooling counteracted by internal heating, or if fast cooling is not active at all. 

In this paper we focus on Joule heating.
Joule heating rates may be sufficiently high to hide the cooling effect of nucleonic direct Urca, as discussed in \cite{Aguilera_2008}. Here we extend those results to include hyperon species and to consider stars with $B_{\textrm{dip}} \lesssim 10^{14}$ G and strong internal fields. Although $B_{\textrm{dip}}$ can be inferred from timing properties, there is no direct method to infer the strength of the internal field, which may decay and keep the star hot via Joule heating. Other internal heating mechanisms are also plausible but are not modelled in this paper, such as vortex creep \citep{Shibazaki_1989, Page_2006} or rotochemical heating \citep{Reisenegger_1995, Hamaguchi_2019} (see \cite{Gonzalez_2010} for a concise review).

\section{Theoretical cooling curves and $L_{\gamma}$ degeneracy}
\label{sec:Observations}

In this section we compare the theoretical cooling rates of some of the initial magnetic configurations reported in Tables \ref{tab:Topology_Table_crust} and \ref{tab:Topology_Table_core} with the available data of isolated, young magnetars with $B_{\textrm{dip}} \gtrsim 10^{14}$ G and of older neutron stars with $t \gtrsim 10^5$ yr and $B_{\textrm{dip}} \lesssim 10^{14}$ G. For magnetars, we use the data corresponding to $16$ objects with ages $\lesssim 10^7$ yr reported in \cite{Vigano_2013}. Typically, the sources have inferred magnetic fields with $B_\textrm{dip} \gtrsim 10^{14}$ G (some $B_\textrm{dip}$ values have been updated\footnote{We refer the reader to the McGill online magnetar catalogue \url{http://www.physics.mcgill.ca/~pulsar/magnetar/main.html} \citep{Olausen_2014}}). For stars with weaker fields, we are mostly interested in ages $\gtrsim 10^4$ yr, and we use the data reported in \cite{Potekhin_2020}.
We consider two typical masses, namely a low-mass model with $M = 1.3 \ M_{\odot}$ (without hyperons in the stellar core) and a high-mass model with $M = 1.8 \ M_{\odot}$ (with hyperons). We use the nucleon and hyperon gap models specified in the previous section. The magneto-thermal evolution of models with $M = 1.8 \ M_{\odot}$ and hyperon cores is studied in detail in Appendix \ref{sec:MT_evol}.

\subsection{Magnetars}

\begin{figure*}
\includegraphics[width=17cm, height = 6.5cm]{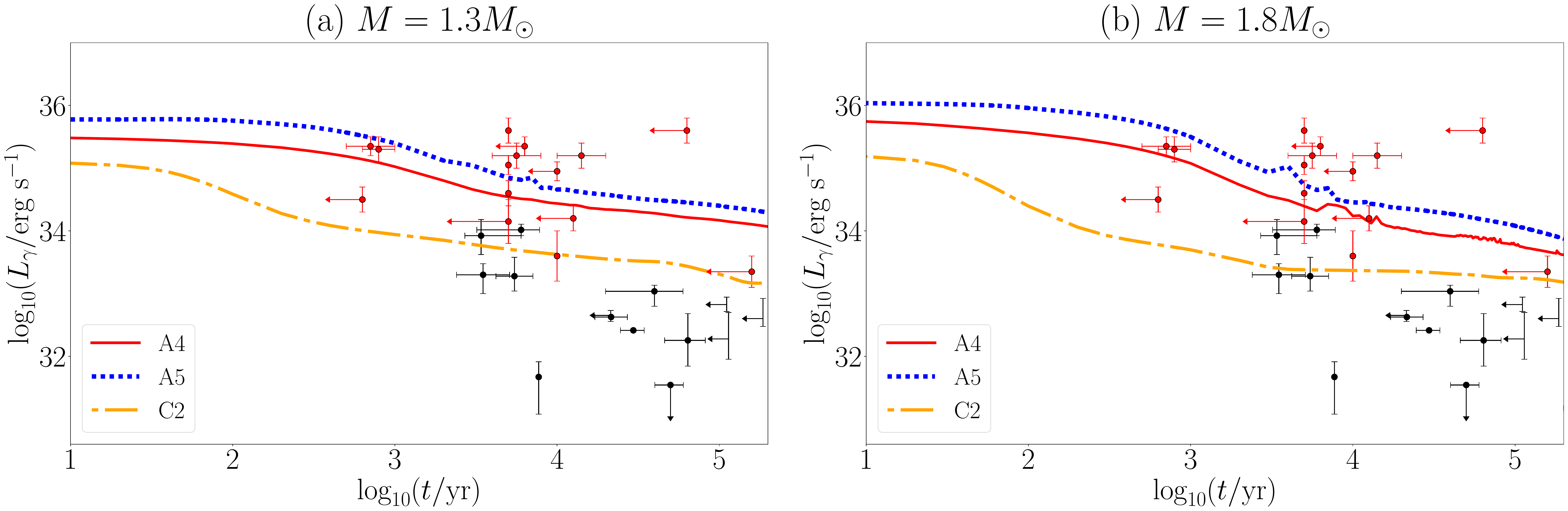}
\caption{Cooling curves for models with $M = 1.3 \ M_{\odot}$ and $M = 1.8 \ M_{\odot}$, with nucleons and hyperons in the superfluid phase. Overlapped are the data points corresponding to magnetars (red dots, \citep{Vigano_2013}) and data corresponding to moderately magnetized stars (black dots, \citep{Potekhin_2020}). \textit{(a)} $M = 1.3 \ M_{\odot}$. \textit{(b)} $M = 1.8 \ M_{\odot}$. The legends report the initial magnetic field configurations (see Tables \ref{tab:Topology_Table_crust} and \ref{tab:Topology_Table_core} for details).}
\label{fig:Cool_curves_magnetars}
\end{figure*}

We first focus on magnetars, which typically have an inferred dipolar poloidal field strengths at the polar surface in the range $10^{14} \ \textrm{G} \lesssim B_{\textrm{dip}} \lesssim 10^{15}$ G and ages $t \lesssim 10^5$ yr.

In Figure \ref{fig:Cool_curves_magnetars} we display the cooling curves corresponding to the A4, A5 and C2 configurations (red, blue and orange curves respectively). Magnetars are represented by red dots; stars with lower $B_{\textrm{dip}}$ are represented by black dots. Figure \ref{fig:Cool_curves_magnetars}(a) studies the model with $M = 1.3 \ M_{\odot}$, which cools mainly via nucleonic direct Urca, and Figure \ref{fig:Cool_curves_magnetars}(b) reports the cooling curves corresponding to the model with $M = 1.8 \ M_{\odot}$, cooling via both nucleonic and hyperonic direct Urca.

In Figure \ref{fig:Cool_curves_magnetars}(a), the A5 configuration (blue, dotted curve) maintains higher $L_{\gamma}$ with respect to the A4 configuration (red, solid curve) and the C2 configuration (orange, dotted-dashed curve). The blue curve matches some of the most luminous sources with $L_{\gamma} \gtrsim 10^{35}$ erg s$^{-1}$. The model maintains $L_{\gamma} \gtrsim 10^{34}$ erg s$^{-1}$ up to $t \approx 2 \times 10^{5}$ yr. The red cooling curve attains lower values of $L_{\gamma}$, and at later times ($t \gtrsim 10^4$ yr) it maintains a similar thermal luminosity to the blue curve ($L_{\gamma} \gtrsim 10^{34}$ erg s$^{-1}$). The C2 configuration matches lower thermal luminosities of both magnetars and stars with lower fields.

In Figure \ref{fig:Cool_curves_magnetars}(b) we study the same magnetic configurations as in Figure \ref{fig:Cool_curves_magnetars}(a), but for a star with $M = 1.8 \ M_{\odot}$ and superfluid hyperons. The blue dotted and red solid lines (A5 and A4 configurations respectively) attain similar thermal luminosities to the corresponding low-mass models in Figure \ref{fig:Cool_curves_magnetars}(a) for $10^3 \ \textrm{yr} \lesssim t \lesssim 10^5$ yr. For $10^5 \ \textrm{yr} \lesssim t \lesssim 2 \times 10^5$ yr, both curves fall below $L_{\gamma} \approx 10^{34}$ erg s$^{-1}$, contrarily to the corresponding curves in Figure \ref{fig:Cool_curves_magnetars}(a). The orange curve (C2 configuration) attains lower $L_{\gamma}$ with respect to the corresponding curve in Figure \ref{fig:Cool_curves_magnetars}(a) for $t \lesssim 10^4$ yr. At later times, the orange curves in Figure \ref{fig:Cool_curves_magnetars}(a) and Figure \ref{fig:Cool_curves_magnetars}(b) reach similar $L_{\gamma}$.

The comparison between the cooling curves displayed in Figure \ref{fig:Cool_curves_magnetars}(a) and Figure \ref{fig:Cool_curves_magnetars}(b) shows that the thermal luminosity observations and age estimates of magnetars can be explained equally by stars cooling mainly via nucleonic direct Urca emission ($M = 1.3 \ M_{\odot}$) and stars cooling mainly via both nucleonic and hyperonic direct Urca emission ($M = 1.8 \ M_{\odot}$). The decay of the strong magnetic field produces sufficient thermal power to decouple the thermal evolution of the crust and the core, and the measured $L_{\gamma}$ of magnetars is dominated by Joule heating in the crust, regardless of the neutrino emission mechanisms active in the core involving hyperons.

\begin{figure*}
\includegraphics[width=17cm, height = 6.5cm]{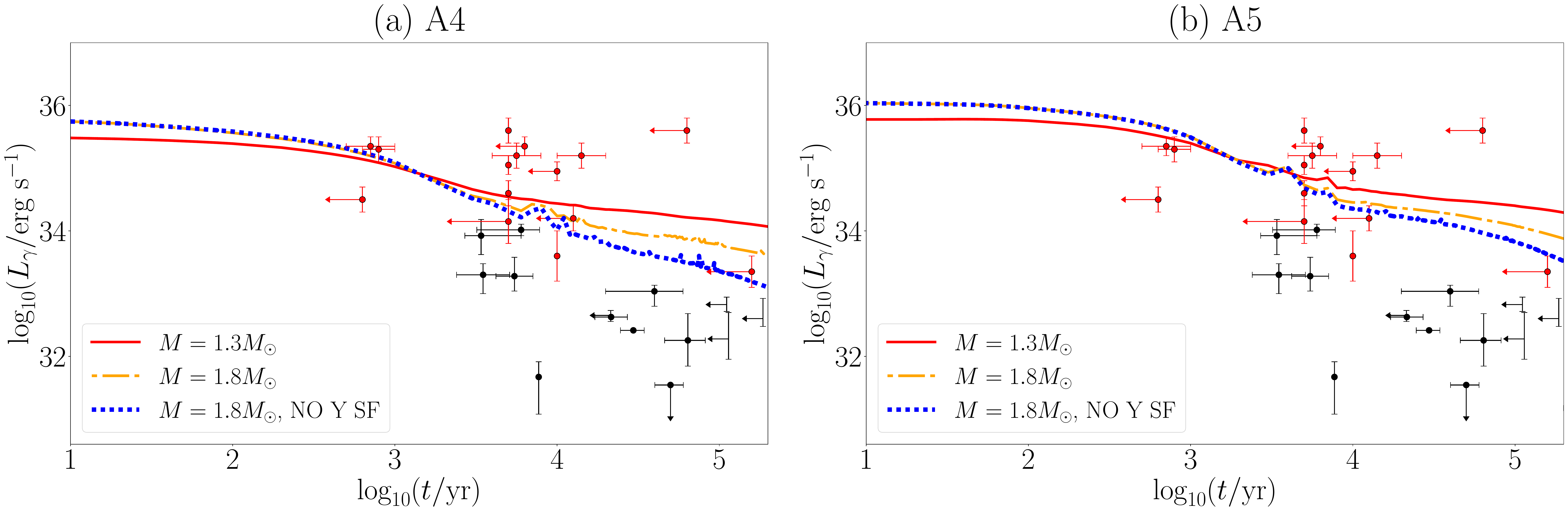}
\caption{Effect of hyperon superfluidity on the cooling curves of models with magnetar-like fields. \textit{(a)} A4 initial magnetic configuration. \textit{(b)} A5 initial magnetic configuration. In both panels, the blue dotted curves ($M = 1.8 \ M_{\odot}$, with hyperons in the normal phase) and orange dotted-dashed curves ($M = 1.8 \ M_{\odot}$, with superfluid hyperons) are degenerate for $t\lesssim 10^4$ yr; the red solid curves ($ M = 1.3 \ M_{\odot}$) and orange curves are similar for $t \lesssim 10^4$ yr. Nucleons are superfluid in both panels.}
\label{fig:Degeneracy_magnetars}
\end{figure*}

Similar conclusions hold if hyperons are not superfluid, and hyperon direct Urca operates without being suppressed by superfluid effects (Figure \ref{fig:Degeneracy_magnetars}). The cooling curves in Figure \ref{fig:Degeneracy_magnetars}(a) correspond to models with the A4 initial magnetic configuration. The red curve shows the case $M = 1.3 \ M_{\odot}$, and the orange dotted-dashed curve the case $M = 1.8 \ M_{\odot}$ with hyperon superfluidity. The blue, dotted curve is obtained for $M = 1.8 \ M_{\odot}$ without hyperon superfluidity. Up to $t \approx 10^4$ yr, the red, orange and blue cooling curves are similar, and are compatible with the $L_{\gamma}$ data of the same magnetars. The orange and blue curves are degenerate up to $t \approx 10^4$ yr. Only at later times Joule heating becomes weaker (e.g. $t\gtrsim 10^4$ yr), and one can distinguish between low-mass and high-mass stars, with or without hyperon superfluidity. Figure \ref{fig:Degeneracy_magnetars}(b) reports a scenario similar to Figure \ref{fig:Degeneracy_magnetars}(a), but for the A5 initial configuration. The red and orange curves ($M = 1.3 \ M_{\odot}$ and $M = 1.8 \ M_{\odot}$ models respectively, the latter including hyperon superfluidity) and the blue curve ($M = 1.8 \ M_{\odot}$, without hyperon superfluidity)  are similar up to $t \approx 2 \times 10^4$ yr. As in Figure \ref{fig:Degeneracy_magnetars}(a), the three cooling curves are distinguishable only for $t \gtrsim 2\times 10^4$ yr, which exceeds most of the age estimates of the magnetar population \citep{Vigano_2013}.

Some magnetar sources lie above the blue curves in Figure \ref{fig:Cool_curves_magnetars}. We remind the reader that in this work we consider magnetized, iron-only outer envelopes. The data points with $10^{35} \lesssim L_{\gamma}/\textrm{erg s}^{-1} \lesssim 10^{36}$  may be explained by invoking accreted envelopes and/or higher initial magnetic fields \citep{Potekhin_2001, Potekhin_2003,Vigano_2013}. Furthermore, the thermal luminosity of these magnetars may be higher because of the presence of small hot spots, produced by the inflow of magnetospheric currents on the stellar surface. The latter process is not modeled in our simulations. We also note that our simulations do not include Joule heating in the highly resistive layer of the outer envelope, which may help to increase the thermal luminosity and match the data of the brightest magnetars. 

Above we consider low-mass and high-mass models obtained with the same EoS. However, there are several more scenarios in which the cooling curves become nearly indistinguishable. Consider for example two models with the same (high) mass, the first obtained with the GM1A EoS (hosting $npe\mu Y$ matter and cooling via nucleonic and hyperonic direct Urca) and the second with a different EoS, for example hosting only $npe\mu$ matter and cooling only via nucleonic direct Urca. If both stars are born with strong magnetic fields, their magneto-thermal evolution can lead to similar observed thermal luminosities, making it hard to infer the presence of hyperons in the stellar core. We also emphasize that our results depend unavoidably on the superfluid model adopted, in particular on the nucleon energy gaps in the core. Smaller nucleon gaps lead to higher nucleon direct Urca emissivity, widening the $L_{\gamma}$ difference between light and heavy models. In this case, stronger Joule heating may be required to obtain the $L_{\gamma}$ degeneracy discussed above.

In summary, thermal luminosity data of magnetars with $L_{\gamma} \gtrsim 10^{34}$ erg s$^{-1}$ are not suitable to infer whether the core contains hyperons or not and hence infer the internal composition. We note also that it is not possible to constrain the properties of hyperon superfluid phases, since $L_\gamma$ is degenerate for stars with cores hosting normal or superfluid hyperons with large energy gaps. Our results show that low-mass models composed of $npe\mu$ matter and high-mass models composed of $npe\mu Y$ matter have a similar magneto-thermal evolution due to crust-core decoupling.

\begin{figure*}
\includegraphics[width=17cm, height = 6.5cm]{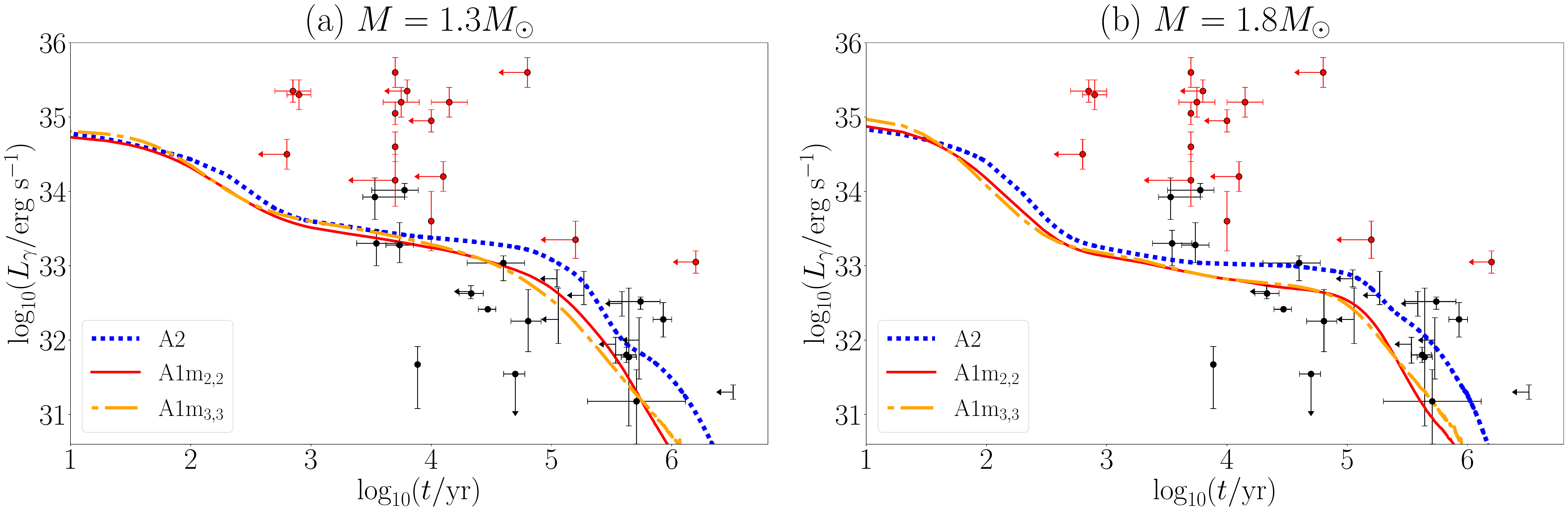}
\caption{Cooling curves for \textit{(a)} $M = 1.3 \ M_{\odot}$ and \textit{(b)} $M = 1.8 \ M_{\odot}$ models with the A2, A1m$_{2,2}$, A1m$_{3,3}$ initial magnetic configurations (see Table \ref{tab:Topology_Table_crust}). As in Figures \ref{fig:Cool_curves_magnetars} and \ref{fig:Degeneracy_magnetars}, the red data points correspond to magnetars (taken from \citet{Vigano_2013}). The black dots corresponds to stars with $B_{\textrm{dip}} \lesssim 10^{14}$ G (taken from \citet{Potekhin_2020}). Nucleon and hyperon species are superfluid.   }
\label{fig:Cool_curves_LowB}
\end{figure*}

\begin{figure*}
\includegraphics[width=17cm, height = 6.5cm]{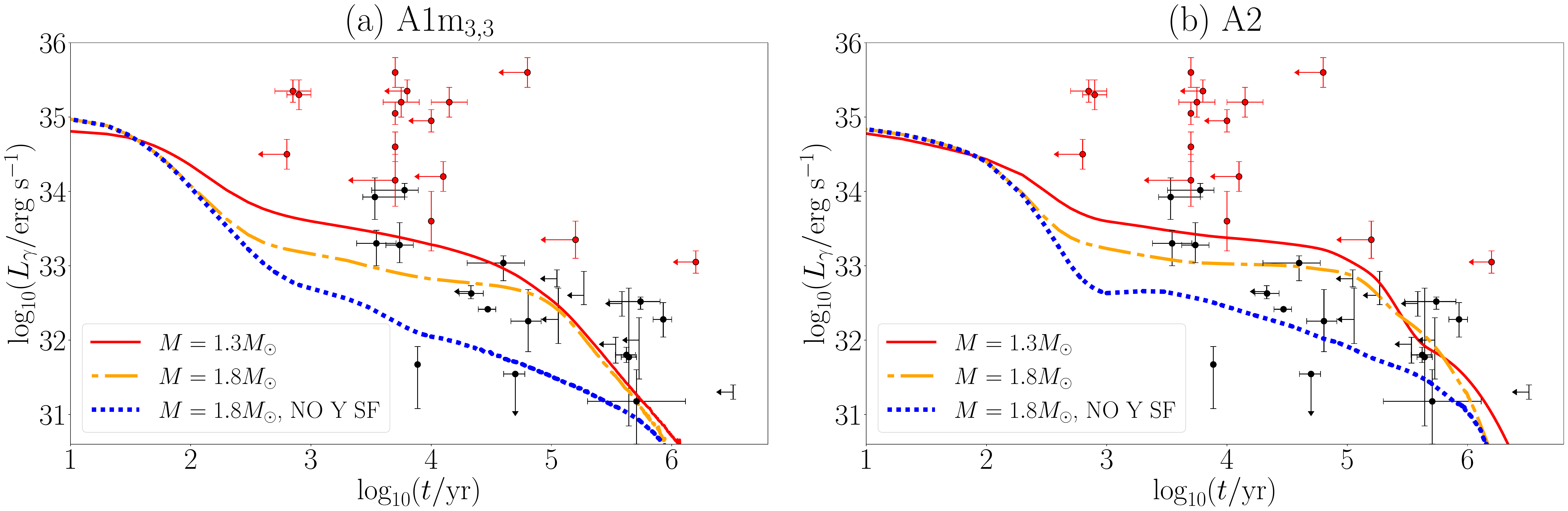}
\caption{Effect of hyperon superfluidity on the cooling curves of stars with $B_{\textrm{dip}} \lesssim 10^{14}$ G. \textit{(a)} A1m$_{3,3}$ initial magnetic configuration. The solid red ($M = 1.3 \ M_{\odot}$) and dotted-dashed orange ($M = 1.8 \ M_{\odot}$ with hyperon superfluidity) cooling curves are similar for $t \gtrsim 10^5$ yr. The blue curve ($M = 1.8 \ M_{\odot}$ without hyperon superfluidity) is clearly distinguishable from the orange and red ones. \textit{(b)} As in panel (a), but for the A2 initial magnetic configuration. The red solid and orange dotted-dashed curves are similar for $t \gtrsim 10^5$ yr. There is no degeneracy between the blue and orange curves, unlike in Figure \ref{fig:Degeneracy_magnetars}. In both panels, nucleons are in the superfluid phase.}
\label{fig:Degeneracy_LowB}
\end{figure*}

\subsection{Low-$B_{\textrm{dip}}$ neutron stars}

Magnetars are only a subset of the observable neutron star population. Timing measurements reveal that most neutron stars have inferred fields satisfying $B_{\textrm{dip}} \lesssim 10^{14}$ G. We study the evolution of such stars in Figure \ref{fig:Cool_curves_LowB}, where we display the cooling curves corresponding to the A2, A1m$_{2,2}$ and A1m$_{3,3}$ configurations (blue dotted, red solid and orange dotted-dashed curves respectively). The superfluid energy gaps are the same as in Figure \ref{fig:Cool_curves_magnetars}.   

Figure \ref{fig:Cool_curves_LowB}(a) reports the case $M = 1.3 \ M_{\odot}$. For $t \lesssim 10^4$ yr, the blue, red and orange lines attain similar thermal luminosities. The A1m$_{2,2}$ and A1m$_{3,3}$ configurations produce almost degenerate cooling curves, which are compatible with X-ray emitting isolated neutron stars (XINS), such as RX J$1856.5$--$3754$ and RX J$1605.3$+$3249$, and ordinary pulsars such as PSR J$0357$+$3205$ \citep{Potekhin_2020}. 

The case $M = 1.8 \ M_{\odot}$ with superfluid hyperons is presented in Figure \ref{fig:Cool_curves_LowB}(b). The blue, red and orange curves correspond to the A2, A1m$_{2,2}$ and A1m$_{3,3}$ initial magnetic configurations respectively. They differ from the curves in panel (a) for $t \lesssim 10^5$ yr, attaining lower values of $L_{\gamma}$. However, at later times they match the same sources as in Figure \ref{fig:Cool_curves_LowB}(a), e.g. RX J$1605.3$+$3249$ and PSR J$0357$+$3205$. It is not possible to distinguish at $t \gtrsim 10^5$ yr between the cooling curves of low-mass stars cooling via nucleonic direct Urca (Figure \ref{fig:Cool_curves_LowB}(a)) and high-mass stars cooling via both nucleonic and hyperonic direct Urca (Figure \ref{fig:Cool_curves_LowB}(b)). The cooling due to the appearance of hyperons is masked by the high Joule heating rate caused by the decay of the unobserved internal field. One may ask why the cooling curves in Figure \ref{fig:Cool_curves_LowB}(a) differ from the corresponding ones in Figure \ref{fig:Cool_curves_LowB}(b) for $t \lesssim 10^5$ yr, and attain similar values of $L_{\gamma}$ for $t \gtrsim 10^5$ yr. The reason is that the thermal power supplied by Joule heating for the A2, A1m$_{2,2}$ and A1m$_{3,3}$ initial configurations is insufficient to counterbalance the power lost due to nucleonic and hyperonic direct Urca emissivity when the star is relatively hot. The crust-core decoupling is incomplete. However, at later times the direct Urca emissivity is weaker due to the lower internal temperature, and Joule heating dominates the thermal evolution of the star. Consequently, the cooling curves of low-mass and high-mass stars are similar for $t \gtrsim 10^5$ yr.

Below we investigate further the consequences of the incomplete crust-core thermal decoupling in stars with $B_{\textrm{dip}} \lesssim 10^{14}$ G. We show that if hyperons are in the normal phase, the cooling curves of high-mass hyperon stars are clearly distinguishable from the curves of stars without hyperons in their core. Figure \ref{fig:Degeneracy_LowB} displays the cooling curves corresponding to the A1m$_{3,3}$ and A2 magnetic configurations. In Figure \ref{fig:Degeneracy_LowB}(a) (A1m$_{3,3}$ initial configuration) the red (solid) and orange (dotted-dashed) lines correspond to models with $M = 1.3 \ M_{\odot}$ and $M = 1.8 \ M_{\odot}$ (the latter assuming that hyperons are superfluid). The blue, dotted line corresponds to a model with $M = 1.8 \ M_{\odot}$, but with hyperons in the normal phase. The blue curve falls below $L_{\gamma} = 10^{33}$ erg s$^{-1}$ already for $t \lesssim 10^3$ yr, and matches the data corresponding to PSR J$0357$+$3205$ for example. There is no degeneracy between the orange and the blue curve. 

Similar results are found in Figure \ref{fig:Degeneracy_LowB}(b) (A2 initial configuration), where the red and orange curves become almost degenerate for $t \gtrsim 10^5$ yr. However, the blue curve ($M = 1.8 \ M_{\odot}$ model without hyperon superfluidity) is clearly distinguishable, attaining $L_{\gamma} \lesssim 10^{32}$ erg s$^{-1}$ for $t \gtrsim 10^5$ yr. 

In summary, we find two trends. If the star is born with a magnetar-like magnetic field with $B_{\textrm{dip}} \gtrsim 10^{14}$ G and/or a strong internal field that stores a large fraction of the total magnetic energy, the crustal temperature is regulated by Joule heating and is almost independent of neutrino cooling in the core, causing crust-core thermal decoupling. Magnetar data can be explained by both low-mass and high-mass models, regardless of the presence of hyperons. If the star has a lower field at birth ($B_{\textrm{dip}} \lesssim 10^{14}$ G) but the internal field stores most of the magnetic energy, the crust-core thermal decoupling is incomplete. In the latter scenario, if hyperons are superfluid, the magneto-thermal evolution is similar for low-mass and high-mass stars for $t \gtrsim 10^5$ yr. This raises the question of how to distinguish between stars that cool via nucleonic and hyperonic direct Urca heated by magnetic field decay, and stars that cool down only via nucleonic direct Urca (or even stars where direct Urca is not active at all), given that the internal field configuration and strength are unknown. On the contrary, if hyperons are not superfluid, the cooling curves are clearly distinguishable also for $t \gtrsim 10^5$ yr. We emphasize that we do not claim that neutron stars with low inferred values of $B_{\textrm{dip}}$ always have strong internal fields, nor that the available data of thermal emitters with low $B_{\textrm{dip}}$ must be interpreted in terms of strong internal heating. Such magnetic configurations may be characteristic of a subset of the neutron star population, rather than a common feature of thermally emitting stars.

\begin{figure*}
\includegraphics[width=17.5cm, height = 17.5cm]{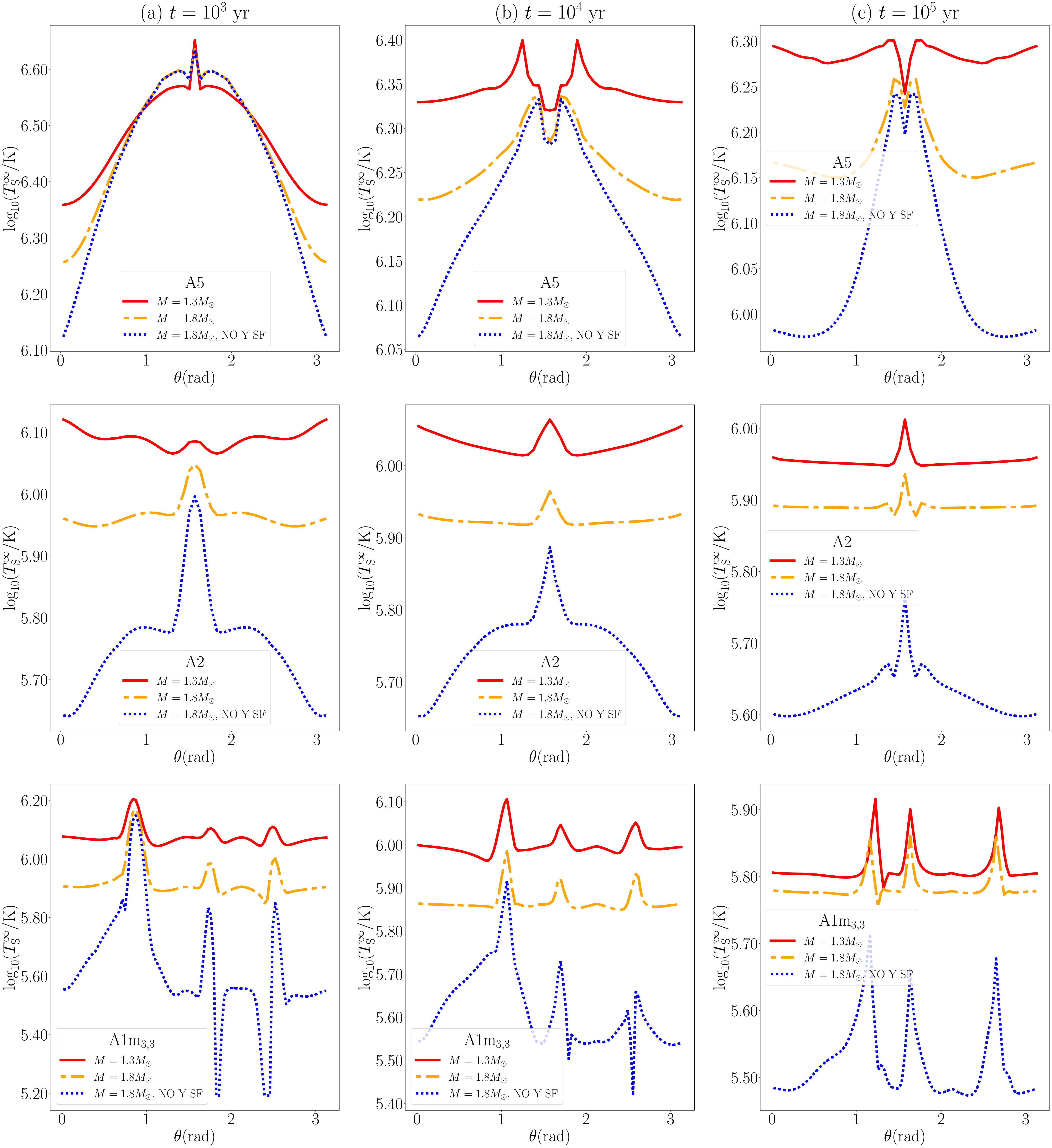}
\caption{Snapshots of the redshifted surface temperature $T^{\infty}_{\textrm{S}}$ versus the colatitude $\theta$, taken at times $t = 10^3, 10^4, 10^5$ yr (left, middle and right columns). The top row reports the evolution corresponding to the A5 configuration; the middle row studies the A2 configuration; the bottom row studies the A1m$_{3,3}$ configuration.}
\label{fig:Surface_temp}
\end{figure*} 

\section{Surface temperature}
\label{sec:Surface_temp}

We now calculate the surface temperature of some of the models reported in Figures \ref{fig:Degeneracy_magnetars} and \ref{fig:Degeneracy_LowB}. We study the similarities between the redshifted surface temperature ($T^{\infty}_{\textrm{S}}$) of low- and high-mass models in Figure \ref{fig:Surface_temp} for the A5, A2 and A1m$_{3,3}$ initial magnetic configurations.

The top row in Figure \ref{fig:Surface_temp} displays snapshots of $T^{\infty}_{\textrm{S}}$ versus the colatitude $\theta$ taken at $t = 10^3, 10^4, 10^5$ yr for the A5 configuration. At $t = 10^3$ yr, the values of $T^{\infty}_{\textrm{S}}$ for the $M = 1.3 \ M_{\odot}$ model (red curve), the $M = 1.8 \ M_{\odot}$ model with hyperons in the superfluid phase (orange curve) and the $M = 1.8 \ M_{\odot}$ model with normal hyperons (blue curve) are similar, being higher at the equator than at the poles. At later times ($t \gtrsim 10^4$ yr), the $M = 1.8 \ M_{\odot}$ model with hyperons in the normal phase cools more quickly due to the higher neutrino emissivity, increasing the difference between the blue and the orange and red curves at the poles. At the equator, the curves remain similar up to $t = 10^5$ yr. In line with the results of $L_{\gamma}$ reported in Figure \ref{fig:Degeneracy_magnetars}, the three models have a similar magneto-thermal evolution, producing almost indistinguishable cooling curves up to $t \approx 2 \times 10^4$ yr, and a similar surface temperature map.

For initial configurations with weaker fields (middle and bottom rows in Figure \ref{fig:Surface_temp}, A2 and A1m$_{3,3}$ configurations respectively), we find the opposite trend for the red and orange curves with respect to the A5 configuration. For example, for the A2 configuration we find that the red and orange curves become increasingly similar as the star cools. This evolution reflects the trend of the cooling curves displayed in Figure \ref{fig:Degeneracy_LowB}: as the internal temperature decreases, the cooling effect of direct Urca in the core weakens, and Joule heating in the crust dominates the thermal evolution and hence $L_{\gamma}$ and $T^{\infty}_{\textrm{S}}$. On the contrary, the blue curve shows that $T^{\infty}_{\textrm{S}}$ remains substantially lower if hyperons are in the normal phase, as Joule heating is not sufficient to control the thermal evolution due to the high emissivity of hyperon direct Urca. The bottom row in Figure \ref{fig:Surface_temp} shows the presence of three hot spots above and below the equator. The surface temperature of light and heavy models becomes similar for $t \gtrsim 10^4$ yr, if hyperons are superfluid. Note that $T^{\infty}_{\textrm{S}}$ is not symmetric with respect to the equator because of the north-south asymmetric magnetic field configuration (see Appendix \ref{sec:MT_evol} for a detailed study of the corresponding magneto-thermal evolution). Contrarily to the A5 configuration, the values of $T^{\infty}_{\textrm{S}}$ for a star with hyperon concentrations in its core are clearly distinguishable from low-mass stars only if hyperons are not superfluid, producing a difference in the thermal luminosity of approximately one order of magnitude up to $t \approx 10^5$ yr (cf. Figure \ref{fig:Degeneracy_LowB}).

\section{Conclusion}

Measurements of $L_\gamma$ as a function of age are one means of probing the composition of neutron star interiors \citep{Yakovlev_2001, Page_2004, Page_2006, Potekhin_2015, Potekhin_2020}, at least in principle. For example, if it is discovered that $L_\gamma$ is lower than predicted theoretically for $npe\mu$ matter, one possible scenario is that accelerated direct Urca cooling caused by hyperons is responsible \citep{Prakash_1992, Haensel_1994, Raduta_2018, Raduta_2019}. In this paper, we show that the situation is more complicated, because Joule heating can mask the cooling effects of direct Urca emission \citep{Aguilera_2008}. Specifically, cooling curves of both low-mass stars without hyperon cores and high-mass stars with hyperon cores can explain thermal luminosity data of magnetars and stars with $B_{\textrm{dip}} \lesssim 10^{14}$ G equally well.

We study the magneto-thermal evolution of hyperon stars with crust-confined and core-extended magnetic field configurations. 
Fields sustained by both crustal and core electric currents produce sufficient Joule heating to explain the observed luminosities of both young magnetars ($B_{\textrm{dip}} \gtrsim 10^{14}$ G and $L_{\gamma} \gtrsim 10^{34}$ erg s$^{-1}$ for $t \lesssim 10^5$ yr) and stars with lower fields ($B_{\textrm{dip}} \lesssim 10^{14}$ G and $L_{\gamma} \lesssim 10^{34}$ erg s$^{-1}$). The internal temperature in the crust is inhomogeneous due to anisotropic electronic transport across and along the field lines and localized Ohmic dissipation \citep{Pons_2019}. If multipolar structures are present, several hot regions appear in the crust, producing inhomogeneous surface temperature maps \citep{Vigano_2013, Dehman_2020}. 

We find that the thermal luminosities of light stars composed of $npe\mu$ matter ($M = 1.3 \ M_{\odot}$) and heavy stars composed of $npe\mu Y$ matter ($M = 1.8 \ M_{\odot}$) become degenerate due to Joule heating. 
Joule heating causes crust-core thermal decoupling in magnetars born with $B_{\textrm{dip}} \gtrsim 10^{14}$ G and/or strong internal fields for $t \lesssim 2 \times 10^{4}$ yr. The cooling effect of hyperon direct Urca is masked by the thermal power generated by the dissipation of electric currents in the crust, and the cooling curves corresponding to models with or without hyperons match the same sources. The comparison between high-mass models with hyperons in the superfluid and normal phases shows that Joule heating is sufficient to counterbalance the losses due to hyperon direct Urca processes, even if the latter are not suppressed by superfluid effects. Consequently, most magnetars may not be suitable candidates to infer information regarding the internal composition of the star, or to constrain hyperon superfluidity. In stars with inferred fields satisfying $B_{\textrm{dip}} \lesssim 10^{14}$ G that harbour strong internal fields, the crust-core thermal decoupling is incomplete. For $t \lesssim 10^5$ yr, the cooling curves of low-mass and high-mass stars (with superfluid hyperons) are distinguishable. At later times however, the thermal power supplied by Joule heating dominates the thermal evolution, and the distinction between low- and high-mass stars lessens. If hyperons are superfluid, it remains an open question whether the core composition can be inferred using $L_{\gamma}$ data, given that the internal field configuration and strength are unknown and Joule heating in the crust may dominate the evolution. Such degeneracy can be broken if hyperons are not superfluid, as Joule heating is unable to supply sufficient thermal power to reduce the cooling effect of hyperon direct Urca, when the latter is not suppressed by superfluid effects. 

We stress that the observational degeneracy discussed in this work concerns only $L_{\gamma}$ and $T^{\infty}_{\textrm{S}}$. In principle, accurate mass and radius measurements of thermal emitters can clearly distinguish between light and heavy stars, but they are currently unavailable. Yet, even with sufficiently accurate mass and radius estimates, inferring the internal composition may still be a difficult task, when so many $npe\mu$ and $npe\mu Y$ EoSs lead to similar macroscopic properties of neutron stars.

We conclude with some cautionary remarks. In this work we focus on stars with strong internal fields, and we show that if certain conditions are met, thermal luminosity data cannot be uniquely interpreted in terms of the internal composition. However, it is not clear whether strong internal fields are ubiquitous across the neutron star population or not; for example, stars with $B_{\textrm{dip}} \lesssim 10^{13}$ G may not necessarily contain strong internal fields. Furthermore, the goal of this paper is not to assess whether hyperons are present or not in neutron stars, but rather to determine whether one key signature of their appearance (i.e. hyperon direct Urca emission) has a \enquote{distinguishable} effect on the cooling curves in the presence of high Joule heating rates. We emphasize that in order to draw definitive conclusions about neutron stars with hyperon cores, several microphysical details (such as the EoS \citep{Schaffner_2002, Stone_2007, Fortin_2015, Raduta_2018, Motta_2019, Motta_2022}, superfluid model or the neutrino emissivity for example) and evolutionary details (e.g. typical initial conditions for isolated and binary neutron stars) must be ascertained more accurately than they are at present. For example, for smaller neutron triplet and proton superfluid energy gaps, nucleon direct Urca emission is stronger, and the Joule heating rate required for the cooling curves of light and heavy stars to become degenerate may be higher than the one calculated in this work. Additionally, in our study we focus on massive stars containing hyperons. Stars with $M \gtrsim 1.6 \ M_{\odot}$ are often found in binary systems \citep{Kiziltan_2013, Alsing_2018}, and may experience accretion at different epochs. Accretion alters the magnetic field configuration, heats the surface and internal layers and modifies the chemical composition of the crust \citep{Payne_2004, Haensel_2008, Priymak_2011, Fantina_2018, Potekhin_2019, Gusakov_crust_2020, Gusakov_2021}. Alternatively, massive stars may be remnants of merger events (provided that the remnant does not collapse into a black hole), and their initial conditions are likely to be more complicated than the standard ones employed in this work and in the literature of neutron star cooling. More realistic initial conditions for massive stars will be studied in future work. 

\section*{Acknowledgements}

We thank Prof. Xavier Vi\~nas and Prof. Mario Centelles for providing the BCPM nuclear energy density functional used to test the in-medium correction factors for the modified Urca emissivity. 
FA is supported by The University of Melbourne through a Melbourne Research Scholarship. AM acknowledges funding from an Australian Research Council Discovery Project grant (DP170103625) and the Australian Research Council Centre of Excellence for Gravitational Wave Discovery (OzGrav) (CE170100004). DV is supported by the European Research Council (ERC) under the European Union’s Horizon 2020 research and innovation programme (ERC Starting Grant "IMAGINE" No. 948582, PI DV). CD is supported by the ERC Consolidator Grant “MAGNESIA” (No. 817661, PI Nanda Rea) and this work has been carried out within the framework of the doctoral program in Physics of the Universitat Aut\`onoma de Barcelona. JAP acknowledges support by the Generalitat Valenciana (PROMETEO/2019/071), AEI grant PGC2018-095984-B-I00 and the Alexander von Humboldt Stiftung through a Humboldt Research Award.

\section*{Data availability}
The data corresponding to the EoSs employed in this work are taken from the Web page \url{http://www.ioffe.ru/astro/NSG/heos/hyp.html}, and are presented in \enquote{Physics input for modelling superfluid neutron stars with hyperon cores} \citep{Gusakov_2014}. 
The thermal luminosity and age data of moderately magnetized neutron stars are reported in the paper \enquote{Thermal luminosities of cooling neutron stars} by \cite{Potekhin_2020} and are accessible at the Web page \url{http://www.ioffe.ru/astro/NSG/thermal/cooldat.html}. The magnetar data are taken from the paper \enquote{Unifying the observational diversity of isolated neutron
stars via magneto-thermal evolution models} by \cite{Vigano_2013}.

\bibliographystyle{mnras}
\bibliography{BIBLIO}

\begin{appendix}

\section{Magneto-thermal evolution}
\label{sec:MT_evol}

\begin{figure*}
\includegraphics[width=17.5cm, height = 14cm]{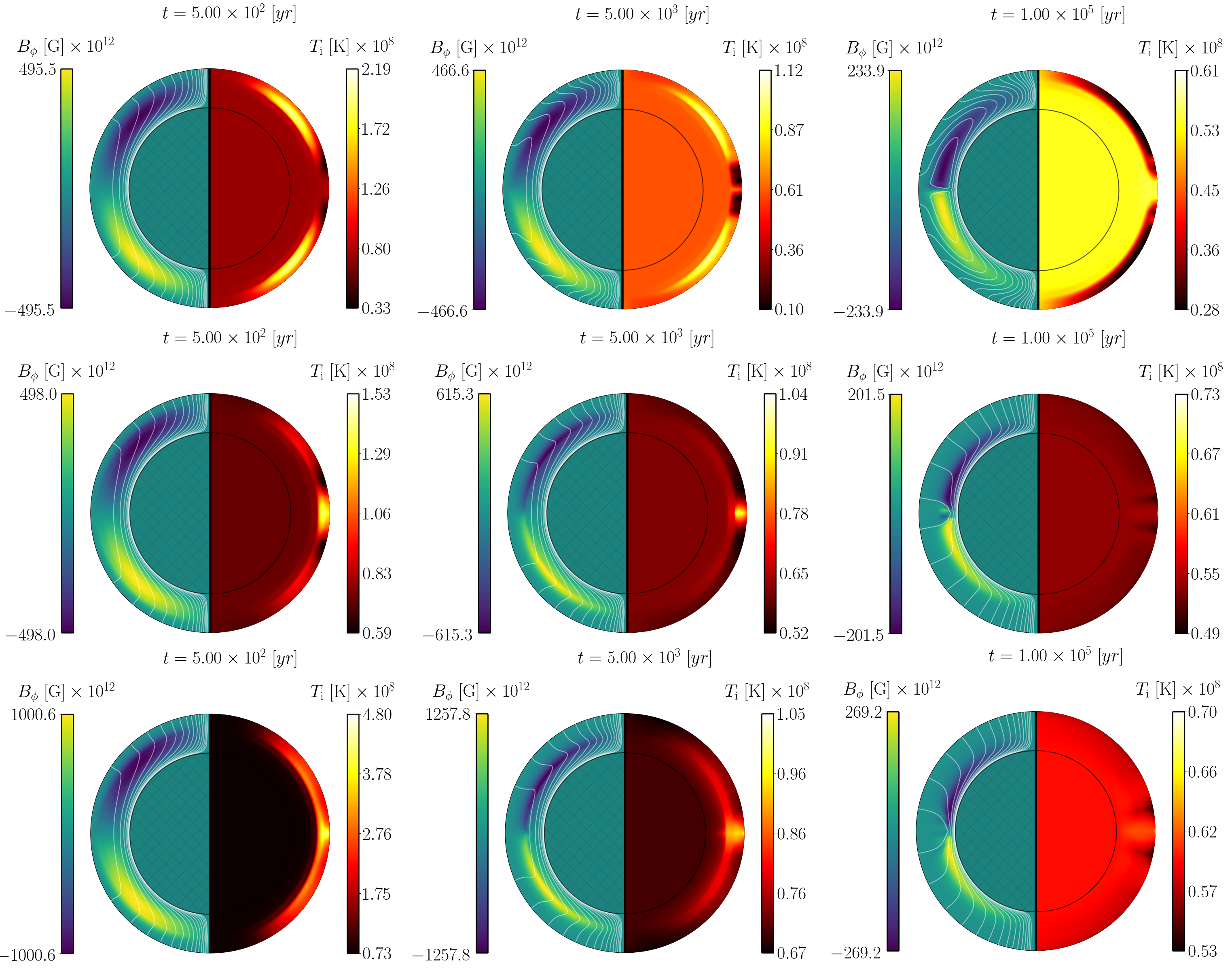}
\caption{Magnetic and thermal evolution of models with $M = 1.8 \ M_{\odot}$ and different initial magnetic configurations. The top, middle and bottom panels correspond to the A1, A2 and A3 configurations respectively (see Table \ref{tab:Topology_Table_crust} for details). The left hemisphere in each polar plot displays the contours of the toroidal magnetic field component $\boldsymbol{B}_{\textrm{tor}} = B_{\phi} \hat{\phi}$ in units of $10^{12}$ G. Overplotted are the poloidal field lines projected on the meridional plane. The map of the internal redshifted temperature $T_{\textrm{i}}$ (in units of $10^8$ K) is displayed in the right hemispheres. The thickness of the crust here and in all the following figures is enlarged by a factor of 8 for visualization purposes.}
\label{fig:Cr_no_multi}
\end{figure*}

\begin{figure*}
\includegraphics[width=17.5cm, height = 14cm]{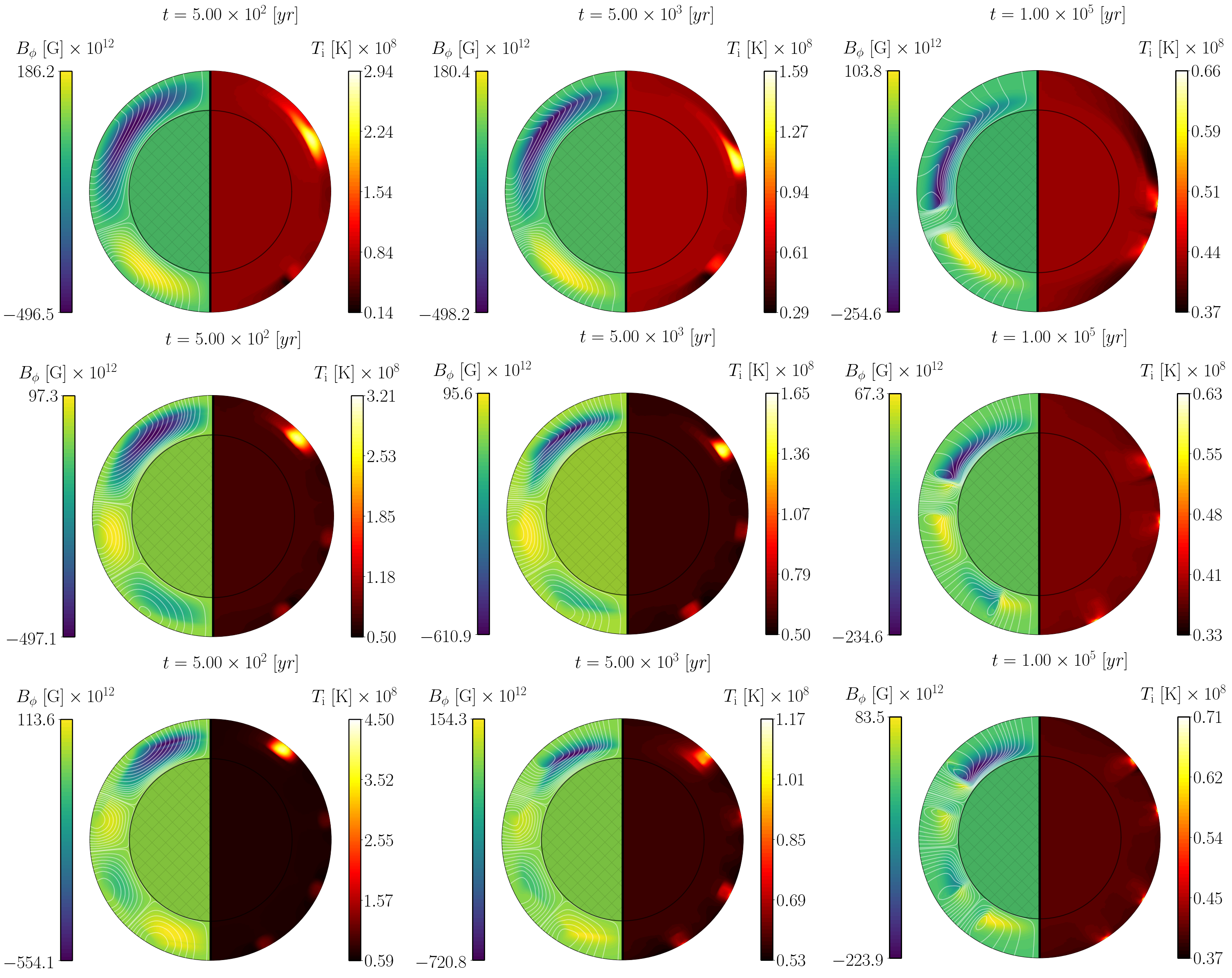}
\caption{As in Figure \ref{fig:Cr_no_multi}, but for the A1m$_{2,2}$, A1m$_{3,3}$ and A1m$_{4,4}$ configurations (top, middle and bottom panels respectively). See Table \ref{tab:Topology_Table_crust} for details.}
\label{fig:Cr_multi}
\end{figure*} 

\begin{figure*}
\includegraphics[width=17.5cm, height = 14cm]{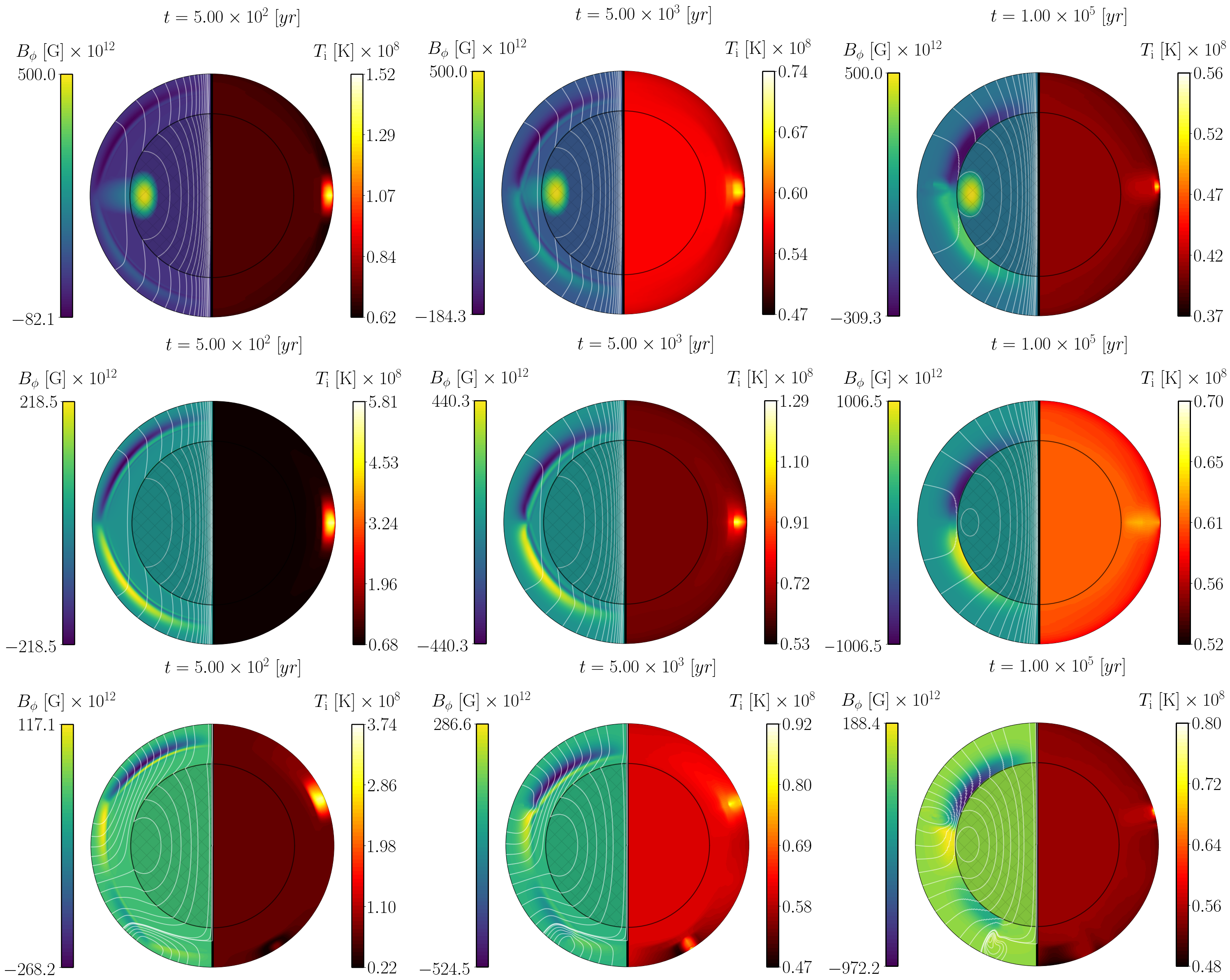}
\caption{Evolution of neutron star models with magnetic fields sustained by core and crustal electric currents. The top panels correspond to the C1 initial configuration, the middle panels to the C2 configuration and the bottom panels to the C1m$_{2,0}$ configuration. Details about the magnetic configurations are specified in Table \ref{tab:Topology_Table_core}. }
\label{fig:Core_crust}
\end{figure*}

In this Appendix we study the magnetic field and internal temperature evolution of a neutron star model with representative mass $M = 1.8 \ M_{\odot}$ hosting nucleons, electrons, muons, $\Lambda$ and $\Xi^-$ hyperons, assuming that nucleon and hyperon species are superfluid. We first consider the case of crust-confined magnetic configurations (listed in Table \ref{tab:Topology_Table_crust}) in Section \ref{sec:crust_confined}. Core-threading magnetic field configurations (listed in Table \ref{tab:Topology_Table_core}) are presented in Section \ref{sec:core_extended}.

\subsection{Crust-confined fields}
\label{sec:crust_confined}

\subsubsection{Dipolar-poloidal, quadrupolar-toroidal fields}

In Figure \ref{fig:Cr_no_multi} we study a star with $M = 1.8 \ M_{\odot}$, whose initial crust-confined magnetic field has dipolar poloidal and quadrupolar toroidal components. In the following, the left hemispheres of the polar plots display the contours of the toroidal component $\boldsymbol{B}_{\textrm{tor}} = B_\phi \hat{\phi}$. Overplotted are the meridional projections of the poloidal field lines. The right hemispheres display the internal redshifted temperature ($T_{\textrm{i}} = T e^{\Phi}$) maps. We allow the $B_\phi$ and $T_{\textrm{i}}$ scales to vary for all the configurations and snapshots, in order to preserve a high level of detail in the magnetic and temperature maps.

The top panels in Figure \ref{fig:Cr_no_multi} show the magneto-thermal evolution of the A1 initial magnetic configuration. At $t = 5 \times 10^2$ yr, the magnetic field configuration (left hemisphere) is almost identical to the one at birth. The corresponding $T_{\textrm{i}}$ map (right hemisphere) shows significant inhomogeneities in the crust (the bottom of the outer envelope is placed at $\approx 10^{10}$ g cm$^{-3}$, corresponding to the outermost boundary in the plots). This effect is related to anisotropic heat transport and localized Joule heating. At $t = 5 \times 10^3$ yr, the poloidal field lines bend above and below the equator. At later times ($t = 10^5$ yr), the poloidal field lines form two large closed meridional loops just below and above the equator. The right hemisphere shows that the equatorial region is hotter than the rest of the star.

The middle panels display the A2 configuration. The evolution of the poloidal field lines is similar to the top panels for $t \lesssim 5 \times 10^3$ yr. At $t = 5 \times 10^3$ yr the maximal values of $B_{\phi}$ are higher with respect to the snapshot at $t = 5 \times 10^2$ yr, revealing a redistribution of magnetic energy between the poloidal and toroidal components due to the Hall term in the induction equation. The right hemisphere shows that at the equator a hotter region forms. By comparing the snapshot at $t = 10^5$ yr in the middle row with the corresponding one for the A1 configuration, one finds two main differences: (1) the closed poloidal loops are absent in the A2 configuration; and (2) the toroidal field in the A2 configuration is more compressed at the bottom of the inner crust, where its dissipation rate is enhanced due to the presence of impurities in the crustal lattice and by pasta phases \citep{Pons_2013, Vigano_2013, Anzuini_2021}. 

The bottom panels report the evolution of the initial A3 configuration. The magnetic field evolution is similar to the one of the A2 configuration (middle panels), but Joule heating is higher than both the A1 and A2 configurations. We note that, as found for the A2 configuration, the snapshot at $t = 10^5$ yr does not display closed poloidal loops. These are present only in the A1 configuration, where the toroidal field stores most of the total magnetic energy.

\subsubsection{Multipolar topologies}

We consider multipolar magnetic fields in Figure \ref{fig:Cr_multi}, where we simulate the magnetic and thermal evolution of neutron stars assuming the presence of two, three or four poloidal and toroidal magnetic multipoles \citep{Dehman_2020}. In reality one may expect multipoles of higher orders; however, currently it is not feasible to evolve such configurations numerically.

As a general feature, the presence of a given number of magnetic multipoles leads to the appearance of an equal number of \enquote{hot regions} inside the stellar crust. The top panels report the evolution of the A1m$_{2,2}$ configuration. As a consequence of the north-south asymmetry in the magnetic field configuration, the temperature distribution is asymmetric with respect to the equator for $t \lesssim 5 \times 10^2$ yr. The northern hemisphere is characterised by a thick, hot layer in the crust, whose temperature is higher than the hot layer below the equator. The electric currents are asymmetric with respect to the equatorial plane; they are more intense in the northern hemisphere and hence produce higher Joule heating rates than in the southern hemisphere. At later times ($t = 5 \times 10^3$ yr) the shorter visible poloidal field line in the northern hemisphere shifts slightly towards the equator, and so does the corresponding hot region in the temperature map. At $t = 10^5$ yr, both the northern and southern hot regions in the right hemisphere extend to denser layers of the crust.

Upon increasing the number of initial multipoles (middle panels, A1m$_{3,3}$ configuration), the magneto-thermal evolution becomes more complex. Three corresponding hot regions appear in the layers beneath the outer envelope in the temperature map (snapshot at $t = 5 \times 10^2$ yr). The hottest one is again located in the northern hemisphere. As in the top panels, the snapshot at $t = 5 \times 10^3$ yr (right hemisphere) shows that the hot regions remain in roughly the same locations shown in the snapshot at $t = 5 \times 10^2$ yr. At $t = 10^5$ yr, while the northern and the equatorial hot regions move closer, the southern one drifts further towards the south pole. As for the A1m$_{2,2}$ configuration, all three hot regions extend to deeper parts of the crust. Interestingly, we note that at $t = 10^5$ yr both positive and negative values of $B_{\phi}$ are enclosed within the plotted poloidal field lines, contrarily to what is found for the A1m$_{2,2}$ configuration.

The bottom panels display the A1m$_{4,4}$ configuration. In general, the hot regions remain approximately in their original locations, and expand progressively towards deeper layers with increasing $t$. At $t = 10^5$ yr, the toroidal field assumes both positive and negative values of $B_{\phi}$ in each region enclosed by the plotted poloidal field lines, similarly to what is seen in the corresponding panel for the A1m$_{3,3}$ configuration.

\subsection{Core-threading fields}
\label{sec:core_extended}

The top panels in Figure \ref{fig:Core_crust} display the snapshots corresponding to the magneto-thermal evolution of the C1 initial configuration. The kinks of the poloidal field lines at the crust-core interface are due to the presence of electric currents both in the crust and the core. At $t = 5 \times 10^2$ yr the magnetic field lines bend in the crust, where the Hall term in the induction equation redistributes magnetic energy between the poloidal and toroidal components, forming a toroidal field and twisting the magnetic field lines. The right hemisphere shows that two equatorially-symmetric colder layers form in the north and southern hemispheres, while a hotter region resides at the equator. At later times ($t = 5 \times 10^3$ yr), the poloidal field lines are warped in the crust due to the Hall term, and $B_\phi$ peaks in denser regions of the crust. The internal, redshifted temperature is inhomogeneous near the equator, and the two cold layers above and below the equator become thicker. At $t = 10^5$ yr, the crustal toroidal field is stronger at the crust-core interface, where the dissipation rate of electric currents is enhanced by the presence of impurities in the crustal lattice and pasta phases. We find that $T_{\textrm{i}}$ is comparable to the crust-confined configurations considered in Section \ref{sec:crust_confined}.

The middle panels in Figure \ref{fig:Core_crust} display the evolution of the C2 initial topology. The difference with respect to the C1 configuration is that the field does not include an initial toroidal component confined to an equatorial torus, and the initial value of $B_{\textrm{dip}}$ is higher. In general, the magneto-thermal evolution of this model is similar to the top panels in Figure \ref{fig:Core_crust}. However, due to the stronger initial field, the star maintains higher temperatures than the C1 configuration.

In the bottom panels we display results for the C1m$_{2,0}$ configuration.
At $t = 5 \times 10^2$ yr, the Hall term in the induction equation generates a strong toroidal field. Given the equatorial asymmetry of the initial crustal field, the temperature map at $t = 5 \times 10^2$ yr is asymmetric as well. A hot layer in the crust below the outer envelope is contained within the shortest plotted poloidal field line visible in the northern hemisphere. Below the equator, a colder layer forms, spanning a smaller region than its northern counterpart. At $t = 5 \times 10^3$ yr, the crustal poloidal field lines bend, while the toroidal field peaks in the middle of the crust. The temperature map shows that the hot regions extend to deeper regions than earlier. At $t = 10^5$ yr the region where $B_{\phi}$ switches from negative to positive values drifts towards the equator and is located at the crust-core interface.

\end{appendix}





\bsp	
\label{lastpage}
\end{document}